\documentstyle[smallhd,12pt]{article}
\def\baselinestretch{1.3}
\begin{document}

\newcommand{\beq}{\begin{equation}}
\newcommand{\eeq}{\end{equation}}
\def\lag{Lagrangian}
\def\ra{\rightarrow}
\def\x{\times}
\newcommand{\RMP}[3]{{\em Rev. Mod. Phys.} {\bf #1}, #2 (19#3)}
\newcommand{\Rep}[3]{{\em Phys. Rep.} {\bf #1}, #2 (19#3)}
\newcommand{\Ann}[3]{{\em Ann. Rev. Nucl. Sci.} {\bf #1}, #2 (19#3)}
\newcommand{\NS}[3]{{\em Nucl. Sci.} {\bf #1}, #2 (19#3)}
\def\etal{{\it et al}}
\def\L{{\cal L}}
\def\etc{{\it etc}}
\def\ie{{\it i.e.}}
\def\mxth{\mathsurround=0pt }
\def\xversim#1#2{\lower2.pt\vbox{\baselineskip0pt \lineskip-.5pt
  \ialign{$\mxth#1\hfil##\hfil$\crcr#2\crcr\sim\crcr}}}
\def\simgr{\mathrel{\mathpalette\xversim >}}
\def\simles{\mathrel{\mathpalette\xversim <}}

\vspace*{6cm}
\begin{center}
{\bf PARTICLE PHYSICS SUMMARY, WHERE ARE WE AND WHERE ARE WE
GOING?}\footnote{Conference Summary,
{\it XXVIIth Rencontres de Moriond on Electroweak Interactions
and Unified Theories},
Les Arcs, France, March 1992.}

\vspace{3ex}

Paul  Langacker \\
University of Pennsylvania \\ Department of Physics \\
Philadelphia, Pennsylvania, USA  19104-6396
  \\  \today, \ \ \ UPR-0512T
\vspace{3ex}

\vspace*{4.5cm}
ABSTRACT
\end{center}

The XXVII$^{\rm th}$ Rencontres de Moriond featured approximately 84
talks on a wide range of topics.  I will try to summarize the
highlights under the hypothesis that $SU_3 \x SU_2 \x U_1$ is correct
to first approximation, concentrating on probes for new physics at
various scales.

\newpage

\section{Introduction}
\begin{itemize}
\item The Standard Model
\item The Great Divide: Out with a Bang or a Whimper
\item The Great Unknown: Electroweak Symmetry Breaking
\item The Scales of New Physics
\begin{itemize}
\item The TeV Scale
\begin{itemize}
\item Supersymmetry
\item New Operators, Particles, Interactions, Mixings
\item The $Z$-Pole, LEP 100, LEP 200
\item The Weak Neutral Current
\item Searches/Parameterizations for New Physics
\item Astrophysics/Cosmology
\item CP Violation
\item CP Violation in the $B$ System
\item Weak Scale Baryogenesis
\end{itemize}
\item 100 TeV: Flavor Changing Neutral Currents
\item $10^{2-19}$ GeV: Neutrino Mass
\item $10^{16-19}$ GeV: The Ultimate Unification
\end{itemize}
\item Conclusion
\end{itemize}

\section{The Standard Model}

I will assume that the standard model is basically correct.  The $SU_2
\x U_1$ (electroweak) sector has been stringently tested by QED,
neutral and charged current interactions, and the properties of the
$W$ and $Z$.  However, it is useful to keep probing to search for
TeV-scale perturbations.  One missing piece is the top quark mass,
$m_t$, which can be thought of as mainly a nuisance parameter in many
of the other tests.  There are two untested aspects of the electroweak
model.  The Higgs sector could well be only a crude approximation.
The non-abelian sector is also not directly tested, but indirect
evidence suggests that it is probably OK.

The strong interaction theory, QCD, is tested in $e^+e^-$ and
hadronic jet production, in deep inelastic scattering, in
$\Upsilon$, decays \etc., which probe
the perturbative structure.  The symmetry
properties of the theory are an excellent confirmation.  However,
there is no smoking gun quantitative test of QCD.  Furthermore, QCD
tests have always been hindered because there is no viable competing
theory.  It is useful to keep probing the strong interaction sector of
the standard model for a number of reasons.

One motivation is to conclusively establish QCD.  There should be
future progress from LEP, HERA, and hadron colliders.  Another probe
involves chiral perturbation theory (ChPT).  Ecker~\cite{ecker:here}
emphasized that predictions for rare kaon decays from the chiral
anomaly can provide a parameter-free test of the underlying chiral
field theories.  The non-anomalous ChPT pieces of the decay amplitudes
can in principle be separated using detailed decay distributions.
Truong~\cite{truong:here} emphasized that the ChPT calculations can be
improved by a unitarization prescription, which incorporates
resonances.

A second motivation is to develop the strong interaction technology
needed for interpreting electroweak and collider experiments.
Linde~\cite{linde:here} reviewed the status of measurements of
$\alpha_s (M_Z)$ from LEP; $\alpha_s$ is crucial for QCD tests (\ie,
to establish the running), for testing the hypothesis of grand
unification, and for the corrections to the hadronic $Z$ width.  The
results of determinations from event shapes, hadronic $Z$ decays, and
hadronic $\tau$ decays are shown in Table~\ref{tab1}.
\begin{table}    \centering
\begin{tabular}{|cc|} \hline
$\alpha_s (M_Z)$ & Source \\ \hline
$0.124 \pm 0.005$ & event shapes \\
$0.133 \pm 0.012$(prelim) & $R_Z = B(Z \ra {\rm had})/B(Z \ra
l\bar{l})$ \\
$0.113 \pm 0.011$(prelim) & $R_\tau = B(\tau \ra {\rm had})/B(\tau \ra
l)$ \\
$0.1155 \pm 0.0024$ & non-LEP \\ \hline
\end{tabular}
\caption{Various measurements of $\alpha_s (M_Z)$.}
\label{tab1}
\end{table}
The most precise value comes from various measurements of the hadronic
event topologies.  The quoted error of $0.005$ is almost all
theoretical and is dominated by uncertainties in the scale at which
the coupling is evaluated.  There was considerable discussion at the
meeting as to whether the error is reliable.  The true uncertainty may
well be larger, at least by a factor of two.  The number $0.124 \pm
0.005$ is actually based on the 1990 LEP data; it involves a new
analysis, known as resummed QCD, in which the order $\alpha^2_s$ terms
are combined with next-to-leading logarithm effects in the theoretical
expressions.  This makes the various determinations more consistent
with each other, but the result is $1 \sigma$ higher than the previous
value $0.117 \pm 0.007$, which was based on the same data.  The
somewhat higher value based on the width for the $Z$ to decay into
hadrons has a larger statistical error but is cleaner theoretically.
The value from hadronic $\tau$ decays\footnote{Dam~\cite{dam:here}
obtained a somewhat different value $0.119 \pm 0.006$ from the same
data.} is theoretically more reliable
than one might at first guess, as was
discussed by Pich\cite{pich:here}.  The measurement actually
determines $\alpha_s (m_\tau)$, which is then extrapolated to $M_Z$.
This also demonstrates the running of $\alpha_s$.  One can average the
LEP determinations to yield $\alpha_s (M_Z) = 0.123 \pm 0.004$, but
since the most precise determination is theory-dominated one should be
careful in using this quantity.  The average value obtained from
non-LEP experiments shown in Table~\ref{tab1} is somewhat smaller than
the new values preferred by the LEP data.  The nominal error is much
smaller, but it is not clear that scale and theory errors have been
fully included.

Another aspect of QCD is developing models for electroweak matrix
elements, jet studies, \etc.  The electroweak matrix elements are
needed to interpret the results of $B\bar{B}$ oscillation and
CP-violation experiments and to extract the CKM matrix.  Jet models are
needed for the interpretation of almost all new physics (and standard
model backgrounds) at colliders.  There are various approaches,
including, ChPT and beyond~\cite{ecker:here,truong:here} and heavy
quark symmetry.  Boucard~\cite{boucard:here} reported on lattice
calculations, including the new result
\beq f_{B_d} \sqrt{B_{B_d}} = 220 \pm 40 \pm (?) MeV \eeq
on the decay constant of the $B_d$.  The second (unknown)
uncertainty is from the use of the quenched approximation.  Lattice
calculations are a very promising direction, but there are still
considerable theoretical uncertainties.

There has been much progress in systematic studies of the $c$ and $b$
spectrum and decays~\cite{summers:here}--\cite{mikenberg:here}.  It is
unlikely that the non-leptonic decays will give indications of new
physics, because the theoretical uncertainties are too large.
However, such studies are needed for the ultimate test of CKM
unitarity, CP violation, \etc.

Yet another role of strong interaction studies is to develop
confidence in calculations of related systems.  Of particular
importance is the possibility of a strongly-coupled spontaneous
symmetry breaking sector of the standard model, the dynamics of which
may be related to that of QCD; such physics can be studied at the SSC
and LHC~\cite{veltman:here,truong:here}.  Green~\cite{green:here}
discussed a string-inspired model and/or realization of QCD, which is
useful as a model, has implications for finite temperature field
theories, and may give new insights into the structure of superstring
theories.

\section{The Great Divide: Out With a Bang or a Whimper}

Although the standard model is extremely successful, it has many
shortcomings.  For example:
\begin{itemize}
\item It is very complicated, with 21 free parameters in the minimal
version including general relativity.
\item It has a complicated gauge structure -- the direct
product of three gauge groups with three distinct couplings.  This
suggests the possibility of some sort of grand unification.
\item The pattern of fermion masses, mixings, and families is a
mystery. Possible solutions involve compositeness or string
theories.
\item There are naturalness problems associated with the Higgs
mass and couplings, suggesting the possibility of supersymmetry or
dynamical symmetry breaking.
\item The strong CP problem requires a severe fine-tuning in the
standard model, suggesting the possibility of a Peccei-Quinn symmetry
or spontaneous CP violation.
\item There is no basic unification of gravity in the standard model,
and it provides no insight into the difficulties of quantum gravity or
of the cosmological constant.  Spontaneous symmetry breaking induces a
vacuum energy (cosmological constant) some ${50}$ orders of magnitude
larger than the experimental limit, requiring a fine-tuned cancellation
between the induced and primordial cosmological constants.  It is not
clear whether superstring theories give any help.  The subject was
reviewed here by Duncan~\cite{duncan}.
\end{itemize}

There are many well-known possibilities for new physics, including
compositeness, grand unification, supersymmetry, or superstring
theories.  It is also possible that there is something completely new
and unexpected, though my own suspicion is that that is unlikely at
the TeV scale.

Let me describe two generic scenarios for new physics, which I refer
to as the great divide, or ``out with a bang or a whimper''.  (Of
course, one can have all sorts of hybrid scenarios in between.)  The
(somewhat discouraging) premise is that progress in particle
physics will eventually draw to a close, hopefully on a time scale of
more than 40 years, presumably with a bang or a whimper.  What I mean
by the whimper scenario is the possibility that nature consists of
onion-like layers of new physics which manifest themselves as one goes
to higher and higher energies.  Examples of this are dynamical
symmetry breaking and compositeness.  The whimper scenario is
intrinsically nonperturbative.  If this is chosen by nature we may, if
we are lucky, unpeel perhaps one more layer at the large hadron
colliders, but it is unlikely that we will ever be able to penetrate
much beyond that.

The contrasting idea is the bang scenario, \ie, that there is, at
least approximately, a desert up to the GUT or Planck scale $(M_P \sim
10^{19} GeV)$.  Such a scenario is perturbative by nature and is the
domain of the elementary Higgs bosons, supersymmetry, grand
unification, and superstring theories.  If nature choose this course
there is some hope of our actually probing all the way to the Planck
scale and to the very early Universe.  Recent successes of the
unification of coupling constants in the supersymmetric extension of
the standard model gives some hint that this may be the correct
approach.

\section{The Great Unknown: Electroweak Symmetry Breaking}

In the bang scenario symmetry breaking is assumed to be due to
elementary Higgs scalars, which are presumably perturbative and weakly
coupled.  In the standard model the Higgs boson mass is
\beq M_H = \sqrt{2 \lambda} v ,\eeq
where $v = 246$~GeV is the weak scale and $\lambda$ is the quartic
Higgs self-interaction.  In principle $\lambda$ could take any value
from $0-\infty$, so there is no real prediction for $M_H$.  However,
$\lambda$ is a running quantity which increases with the scale $\mu$.
In order for the theory to make sense $\lambda$ must remain finite
within the domain of validity of the theory.  This implies the upper
limits
\beq M_H \leq \left\{ \begin{array}{ccc} 200\ GeV\; & , &
{\rm \;theory \;\;valid\;\; to\;\;}
M_P \\ 600\ GeV\; & , &
{\rm theory\;\; valid\;\; to \;}2M_H \end{array} \right. .\eeq
These triviality limits can be justified by lattice calculations,
independent of perturbation theory.

Another problem with an elementary Higgs field is the quadratic
divergence in the Higgs mass.  One finds
\beq M_H^2 \sim M_H^{0\;2} + O(\Lambda^2), \eeq
where the first term represents the bare (lowest-order) mass.  The
second term represents the loop corrections, with $\Lambda$ the scale
of new physics which presumably cuts off the quadratically-divergent
integrals.  Since $M_H$ must be of the order of the electroweak scale
there must be a fine-tuned cancellation between the two terms if $M_H
\ll \Lambda$.  This suggests one of two possibilities: (a)
supersymmetry, in which there are elementary Higgs fields but there
are cancellations between fermion and boson contributions to the
self-energy, eliminating the quadratic divergence; or (b) dynamical
symmetry breaking (DSB), in which there are no elementary scalar
fields and
loop integrals are cut off at the compositeness scale.  There are no
realistic models for dynamical symmetry breaking.

The LEP experiments have excluded a light
standard model Higgs.  The most recent results are
\beq M_H > 53.0, \;\; 47.0, \;\; 52.3, \;\; 51.0 \;GeV\eeq
from ALEPH, DELPHI, L3, and OPAL respectively~\cite{sherwood}.  A
naive combining of these results yields a lower limit of 59.2~GeV, but
strong caveats against this procedure were given at the meeting
because the estimates of backgrounds were based on the limit of $O(50
\;GeV)$.  Future prospects were reviewed by
Janot~\cite{janot,kurihara}.  The upper limit from LEP 100 should
ultimately be $O(60 \;GeV)$.  Above this the Higgs production process
$ Z \ra Z^*H$ will be hidden by an irreducible four-fermion background
in which the $Z$ decays into two fermions, one of which radiates a
virtual photon which decays into two more fermions.

At LEP 200  one will have a sensitivity to
\beq \begin{array}{llrr}
80\; GeV, & {\rm for} & \sqrt{s} = 175 \;GeV, & \L = 150\; pb^{-1} \\
93 \;GeV, & {\rm for} & \sqrt{s} = 190 \;GeV, & \L = 500 \;pb^{-1} \\
130 \;GeV, & {\rm for}& \sqrt{s} = 240 \;GeV, & \L = 500 \;pb^{-1}
\end{array} \eeq
through the virtual $Z$ decay $Z^* \ra ZH$.  The various possiblities
refer to the fact that the energy and luminosity of LEP 200 have never
been well defined.  A total energy of 240~GeV is the maximum that is
possible, and the lower energies are those that are more frequently
discussed.  The process $WW \ra H$ would dominate at a possible NLC,
allowing sensitivity to $M_H = 200\; GeV$ for $\sqrt{s} = 500\; GeV$
and $\L = 10 fb^{-1}$.

Pauss~\cite{pauss} discussed the possibilities of Higgs detection at
the hadron colliders LHC and SSC.  Gluon-gluon fusion $gg \ra H$
should dominate the production for $M_H \leq 700 \;GeV$.  All of the
planned detectors are good for the decay $H \ra ZZ \ra 4l$, which
occurs for $M_H > 2 M_Z$.  Things are more difficult in the
intermediate range below $2M_Z$ but above the LEP range.  The decay $H
\ra 2 \gamma$ may be possible to observe, but more study is necessary.

The situation is more complicated in the minimal supersymmetric
standard model (MSSM)\footnote{Things are even
more complicated in nonminimal
models, but there has been relatively little study.}.  One has two Higgs
scalars $h,H$; one pseudo-scalar $A$; and a pair of charged Higgs
scalars $H^{\pm}$.  At tree-level one expects $m_h < M_Z$, while $H,\,
A, \, H^{\pm}$ can be heavier.  A number of authors have recently
shown that loop corrections may be quite significant if the top quark
mass is large, because there are terms quartic in $m_t$.  This was
discussed by Brignole~\cite{brignole}, who showed that a diagrammatic
calculation confirms previous calculations based on the effective
potential.  A typical result is $m_h \leq 130\; GeV$ for $m_t <
180\;GeV, \, m_{\tilde{q}} < 1\; TeV$.  One usually has $M_{H, \; A,
\; H}^{\pm} \gg m_h$, in which case the $h$ acts like a light standard
model Higgs.  However, there are some regions of parameter space in
which there are relatively light $H, \; A, \; H^{\pm}$.  Then the
signatures at LEP-type energies are more complicated: the decay $Z \ra
Zh$ has an amplitude proportional to $\sin (\beta- \alpha)$, while $Z
\ra hA$ is proportional to $\cos (\beta - \alpha)$, where $\tan \beta$
is the ratio of vacuum expectation values of the two Higgs doublets in
the theory and $\alpha$ is the mixing angle between the two scalars.
Current LEP limits exclude light particles in the range $(M_A, \, m_h)
\leq 40\;GeV$~\cite{sherwood}.

An NLC with $\sqrt{s} = 500$ and $\L= \; 10 fb^{-1}$ would be ideally
suited for studying the MSSM Higgs sector~\cite{janot}.  It would
cover all of the parameter space and would either: (a) find a light
standard model-like $h$; (b) observe the decays $hZ + HA$ or $HZ +hA$;
or (c) rule out the MSSM for all acceptable values of $m_t, \;
m_{\tilde{q}}$.  LEP~200 with $\sqrt{s} = 190$ could not contribute
significantly here, but the higher energy version with $\sqrt{s} =
240$ would cover the part of parameter space corresponding to $m_h <
125\; GeV$.  Most but not all of the range for the MSSM would be
acccessible at hadron colliders \cite{pauss}.  Small regions of
parameter space would not be covered, and there is no claim of a ``no
lose'' theorem.

The limits on charged Higgs particles, such as $t \ra b H^+$at $UA(2)$
\cite{tsemelis} and $H^+,\; H^{++}$ at LEP \cite{sherwood} were
reported.  At present $M_{H^+} \geq 40\;GeV$.

The whimper scenario is characterized by heavy nonperturbative Higgs
fields, technicolor, extended technicolor, composite Higgs bosons,
\etc.  One of the best probes is $W_LW_L \ra W_L W_L$ at hadron
colliders.  For example, there may be bound state vectors or scalars
associated with the nonperturbative physics.
Veltman~\cite{veltman:here} discussed the prospects for studying the
nature of heavy physics by measuring $W_LW_L \ra W_LW_L$ below the TeV
scale and then predicting the results at high energies.
Casalbuoni~\cite{casalbuoni} described the consequences of a BESS
(Breaking Electroweak Symmetry Strongly) model involving composite
gauge bosons and the constraints placed on them by the LEP data.
Truong~\cite{truong:here} emphasized that one should not trust the
predictions of dynamical schemes for the TeV scale unless one can
reliably calculate $\pi \pi$ scattering at QCD from first principles,
and discussed the importance of including unitarity in the models.
Zinn-Justin~\cite{zinnjustin} reviewed aspects of $t \bar{t}$
condensation.  He emphasized that the standard model with $M_H \sim
2m_t$ large is mathematically equivalent to a model without an
elementary Higgs but with a $t \bar{t}$ condensate.
Fritzsch~\cite{fritzsch} discussed other possibilities associated with
a condensate of fourth family fermions in the 1 -- 5 TeV range.  The
mixing of these with the other families leads to interesting effects
such as flavor changing neutral currents (FCNC) and violations of
$V-A$ and universality, including such rare decays as $t \ra c Z, \;
\mu \ra 3e, \; b \ra sg$, and induced right-handed currents such as
$t_R \ra b_R$.  However, the underlying dynamical mechanisms for
generating the mixing and inducing $SU_2$~breaking are rather vague.

Most of the possibilities for a strongly coupled symmetry-breaking
sector are best probed at hadron colliders rather than precision
experiments.  Maiani~\cite{maiani} described some technically related
work on the $m_t \ra \infty$ limit of the standard model.

\section{The Scales of New Physics}

\subsection{The TeV Scale}
\subsubsection{Supersymmetry}

Most of the meeting was devoted to searches for new physics at the TeV
scale.  One major possibility is supersymmetry.  Whether or not there
is supersymmetry in nature is, along with the symmetry breaking
mechanism, the crucial question as we approach the great divide.  As
has already been described the SUSY Higgs sector has implications for
LEP, a possible NLC, and hadron colliders.  Another probe is to look
for the superpartners.  They have little direct effect on precision
observables, except possibly large $m_{\tilde{t}} - m_{\tilde{b}}$
splittings.  However, they indirectly affect the possible unification
of coupling constants, which is different in supersymmetric models due
to the contribution of the superpartners to the running.  However, the
issue will ultimately be settled by searches for the direct production
of the superpartners at the SSC, LHC, and possible $e^+e^-$ colliders.

\subsubsection{New Operators, Particles, Interactions, Mixings}

Precision experiments are useful for searching for many types of new
operators, particles, mixings, and interactions.  These by themselves
may not solve the problems of the standard model.  Rather, they are
remnants of new physics that occurs at higher scales.  As was
emphasized by de~Rujula~\cite{rujula}, any such new physics should be
gauge invariant, or it will undermine the successes of the standard
electroweak model.

Such remnants can be searched for directly at hadron colliders, such
as the $Z'$ search by CDF~\cite{maeshima:here}.  However, much of the
effort has been in indirect searches through precision tests --
including QED, the weak charged and neutral currents, and the
properties of the $Z$ and $W$ -- and in astrophysics and cosmology.
Treille~\cite{treille} gave an overview of the precision tests.  For
example, there is a new QED measurement of the anomalous muon magnetic
moment, $g_\mu -2$, being constructed at Brookhaven; this will improve
the present value by a factor of 20, bringing the precision down to
the level of the electroweak effects.  There is need, however, to have
improved measurements of the low energy cross section for $e^+e^-
\ra$~hadrons to reduce uncertainties from the hadronic component of
the vacuum polarization.  These are also the major theoretical
uncertainty in the $M_Z \leftrightarrow \sin^2 \theta_W$ relation.

There are many precise searches for new physics in the weak charged
current sector, including $\beta, \; \mu, \; \tau, \; K, \; c$, and $b$
decays, as well as tests of the CKM matrix and universality.  These are
especially sensitive to new $W_R$ bosons which couple to right-handed
currents, to mixings between exotic and ordinary fermions, and to a
possible fourth fermion family.  An interesting new result from
TRIUMF~\cite{numao} is
\beq R = \frac{B (\pi \ra e \nu + e \nu \gamma)}{B(\pi \ra \mu \nu +
\mu \nu \gamma)} = (1.2265 \pm 0.0056) \x 10^{-4},\label{61792:7}\eeq
compared with the previous value of $(1.218 \pm 0.014) \x 10^{-4}$.
$R$ probes $e \mu$ universality; one expects $(1.234 \pm 0.001) \x
10^{-4}$ in the standard model, in excellent agreement with
(\ref{61792:7}).  From the new TRIUMF result one extracts the ratio
\beq f_e/f_\mu = 0.9970 \pm 0.0023 \eeq
of the effective $e$ and $\mu$ interaction strengths, in good
agreement with universality.  This is sensitive, for example, to
certain types of leptoquarks
with mass up to 200~TeV, as well as to mixings
between ordinary and exotic fermions.

There was considerable discussion of $\tau$ physics, including $\tau$
physics at LEP~\cite{dam:here}, a mini-review \cite{pich:here}, recent
ALEPH results on $\tau_\tau$~\cite{wiedenmann}, $\tau \ra KX$ at
PEP\cite{ronan}, and on the prospects for a $\tau$-charm
factory~\cite{gonzales}.  An important result is that $\tau \ra
\nu_\tau +$~hadrons is a clean measurement of $\alpha_s
(m_\tau)$~\cite{pich:here} despite the low energy scale.  Also,
lepton universality is well tested by the LEP experiments: the $Z \tau
\tau, \; Zee$, and $Z\mu\mu$ couplings are all equal within the small
uncertainties.  The $\tau$ polarization has been measured by the LEP
experiments, $A_\tau = 0.140 \pm 0.024$~\cite{nash}, in agreement with
the expected $0.136 \pm 0.007$.  There has been a new measurement of
the Michel parameters for $\tau$ decay at ARGUS~\cite{mcfarlane:here},
yielding
\beq \begin{array}{lcl}
\tau \ra e \nu_\tau \nu_{e} & , & \rho = 0.78 \pm 0.05 \\
\tau \ra \mu \nu_\tau \nu_{\mu} & , & \rho = 0.72 \pm 0.08.
\end{array} \label{61792:9}\eeq
$V-A$ for $\tau \ra \nu_\tau$ predicts 3/4, while $V+A$ implies 0.
Thus (\ref{61792:9}) establishes that $V-A$ is correct for the $\tau$
interactions and that the third lepton family is a left-handed doublet
like the other families.  These tests would be much more precise at a
$\tau$-charm factory.

One still has the limit $m_{\nu_\tau} < 35 MeV$ on the $\nu_\tau$ mass
from ARGUS.  However, there is a recent theoretical argument from
nucleosynthesis that the range $(few - 25)\; MeV$ for the $\nu_\tau$
mass is probably excluded~\cite{turner}; it is therefore important
that the laboratory limits be improved to eliminate the small window
above 25~MeV.

There have been important new measurements of $\tau$ decays from
LEP~\cite{dam:here}.  ALEPH reports higher branching ratios $B(\tau
\ra 3 \pi \nu, \; \pi 2 \pi^0 \nu)$ than the world average, and
slightly lower 1-prong rates.  This appears to eliminate the 1 prong
problem, at least as far as the LEP data is concerned.  The other
famous problem concerning the $\tau$ decays is still present.  This is
the fact that given any two of the three quantities $B(\tau \ra l \nu
\bar{\nu}), \; \tau_\tau,  $
and $m_\tau$ one can predict the third.  For
some time there has been an inconsistency, which can be characterized
by a somewhat lower effective coupling of the $\tau$ to the weak
current than expected in the standard model~\cite{dam:here}
\beq \begin{array}{ll}
{\rm non-LEP}:  &      G_\tau/G_\mu = 0.972 \pm 0.015 \\
     {\rm LEP} &      {0.977 \pm 0.012} \\
{\rm combined} &      {0.975 \pm 0.010}.   \end{array}
\label{61892:10}\eeq

The new and more precise LEP results still show a discrepancy at
$2.5 \sigma$.  If this holds up it could be an indication that the
$\tau$ neutrino is a mixture $\nu_\tau = \sin \theta \nu_1 +
\cos \theta \nu_2$ of a light
component $\nu_1$ and a neutrino $\nu_2$ that is too heavy to be
produced in the decay.\footnote{An alternate model for explaining the
effects based on universality violation was described by
Ma~\protect\cite{ma}.} The anomaly could be accounted for by $\cos
\theta \sim 0.975\pm 0.010$.  If the extra neutrino were in a fourth
family it would have to be heavier than $M_Z/2$.  If it were sterile
than it would also affect the invisible width of the $Z$, leading to
an effective number of neutrinos $2 + \cos^4 \theta \sim 2.90 \pm
0.05$, which is somewhat low compared to the LEP value \cite{nash}
$N_\nu = 3.04 \pm 0.04$.  However, there is some indication that the
problem may be disappearing.  The ARGUS group has reported a
preliminary value of $m_\tau$ some 8~MeV lower than the world average.
This is reinforced by preliminary results from Beijing of a lower
$m_\tau$ mass.  If $m_\tau$ were lowered by some 11 -- 17 MeV the
discrepancy would disappear.\footnote{The Beijing group subsequently
announced the preliminary value $m_\tau = 1777 \pm 1\;MeV$
\protect\cite{i1}, considerably lower than the old average of
$1284.1^{+2.7}_{-3.6}\;MeV$.  This raises the values of $G_\tau/G_\mu$
in (\protect\ref{61892:10}) by $0.010$, reducing the discrepancy to
$1.5\sigma$.}

\subsubsection{The $Z$-Pole, LEP 100, LEP 200}

LEP ran very well in 1991, with an integrated luminosity of $17000\;
nb^{-1}$, twice that of 1990~\cite{koutchouk}.  They achieved a
transverse polarization of around 10\%.  This should soon allow a
determination of the $Z$ mass to $\Delta M_Z \ll 20 MeV$ by the method
of resonant depolarization.  However, some small systematic problems
have come up which still have to be worked out.  In particular, the
alignment of $RF$ cavities leads to a shift $\Delta E_{CM} \sim 16 \pm
4 MeV$ of the energies of the OPAL and L3 regions compared to those of
ALEPH and DELPHI.  An interesting effect is that it is believed that
the tidal forces of the moon change the size of the ring by $\Delta r
/r \sim 3 \x 10^{-8}$, leading to a shift of $\Delta E_{CM} \sim 8
MeV$ in the LEP energy.\footnote{This is the first experiment in which
all four forces play a significant role!} This will double in the next
run due to a reconfiguration of the magnets.  These effects can
probably be brought under control, but for the time being have delayed
a more accurate determination of $M_Z$.

The electroweak physics program at LEP was reviewed by Nash
\cite{nash}, and the implications of the experiments were described by
Altarelli \cite{altarelli}.  The hadronic charge asymmetry and the $b$
lifetime and width were discussed in other talks
\cite{kreuner,brandl,kroll:here}.  In the 1989 and 1990 runs the four
experiments accumulated a total of 585K hadrons and 63K leptons, while
in 1991 the totals were 1114K and 118K.  There is still a 20 MeV
uncertainty in the LEP energy, which is the dominant uncertainty in
$\Delta M_Z$.  As discussed above it is hoped\footnote{$M_Z$ is
already measured much more precisely than other $Z$-pole observables,
so an improved value is pretty but not urgent.} that this will soon be
reduced to $\ll 20$~MeV.  There is a point-to-point energy uncertainty
of 10 MeV, which leads to an uncertainty $\Delta \Gamma_Z \sim 5$~MeV
in the $Z$ width.  There is an experimental uncertainty in the
luminosity $\Delta \L/\L$ of 0.5\%, and a common theoretical
uncertainty of 0.3\%, which leads to systematic effects in
$\sigma^h_0, \; \Gamma_{l\bar{l}}, \; \Gamma_{\rm had}$ and
$\Gamma_{\rm inv}$.

The principle electroweak results from LEP from the 1990 and 1991 runs
are shown in Table~\ref{tab2}, as well as the predictions of the
standard model for the global best fit value $m_t =
151^{+21}_{-23}$~GeV and 50~GeV$< M_H < 1000$~GeV.
\begin{table}   \centering
\begin{tabular}{|l|c|c|c|} \hline
Quantity & 1990 & 1991 & standard model \\ \hline
$M_Z$(GeV)  & $91.175 \pm 0.021$  &  -- -- & input \\
$\Gamma_Z$(GeV) &  $2.487 \pm 0.010$  &  $2.499 \pm 0.0075$ &  $2.494
\pm 0.002 \pm 0.006 \pm [0.006]$ \\
$\Gamma_{l\bar{l}}$(MeV)  & $83.2 \pm 0.4$  & $83.52 \pm 0.33$  & $83.7
\pm 0.1 \pm 0.2$ \\
$\Gamma_{\rm had}$(MeV) & $1740 \pm 9$ & $1742 \pm 8$ & $1743
\pm 2 \pm 4 \pm [6]$ \\
$\Gamma_{b \bar{b}}/\Gamma_{\rm had}$ & $0.217 \pm 0.010$ &  & $0.216
\pm 0 \pm 0.001$ \\
$A_{FB}\, {(\mu)}$ & $0.0163 \pm 0.0036$ & $0.0176 \pm 0.0029$ & $
0.0155 \pm 0.0006 \pm 0.0012$ \\
$A_{\rm pol}\, {(\tau)}$ & $0.134 \pm 0.035 $ & $ 0.140 \pm 0.024$ &
$0.136 \pm 0.003 \pm 0.006$ \\
$A_{FB}\, {(b)}$ & $0.126 \pm 0.022$ & $ 0.094 \pm 0.014$ & $0.092 \pm
0.002 \pm 0.004$ \\
$R= \Gamma_{\rm had}/\Gamma_{l \bar{l}}$ & $ 20.92 \pm 0.11$ & $20.86
\pm 0.10$ & $20.82 \pm 0.01 \pm 0.01 \pm [0.07]$ \\
$\sigma_o^h (nb)$ & $41.36 \pm 0.23 $ & $41.13 \pm 0.20$ & $41.41 \pm
0.02 \pm 0.02 \pm [0.06] $ \\
$N_\nu$ & $2.99 \pm 0.05$ & $3.04 \pm 0.04$ & $3$ \\
$g^2_A$ & $0.2492 \pm 0.0012$ & $0.2500 \pm 0.0010$ & $0.2513 \pm
0.0002 \pm 0.0004$ \\
$g_V^2$ & $0.0012 \pm 0.0003$ & $0.00131 \pm 0.00024$ & $0.0011 \pm
0.0001 \pm 0.0001$ \\
$\bar{s}^2_W(A_{FB}\, {(q)})$ & $0.2310 \pm 0.0035$ & $0.2316 \pm
0.0032$ & $0.2325 \pm 0.0004 \pm 0.0007 \pm ?$ \\ \hline
\end{tabular}
\caption{Electroweak results from LEP.}
\label{tab2}
\end{table}
The 1991 column includes the earlier data, and many of the results are
preliminary.  The leptonic width assumes $e, \; \mu, \;
\tau$~universality, which is well established by the individual
partial widths.  The vector coupling $g_V^2$ is mainly determined from
$A_{FB} \, {(\mu)}$.  One notable change from 1990 is that the
forward-backward asymmetry into $b$ quarks, $A_{FB}\, {(b)}$, has
decreased somewhat, into excellent agreement with the standard model.
The previous high value had pulled up the extracted value of $m_t$
considerably.  The quantity $\bar{s}_W^2$ is from the hadronic charge
asymmetry.  The last column shows the standard model predictions in
terms of $M_Z$ \cite{langacker}.  The first uncertainty is from $M_Z$
and $\Delta r$, the second is from $m_t$ and $M_H$, and the third (in
square brackets) is a QCD uncertainty assuming $\alpha_s = 0.124 \pm
0.010$, which is extracted from the LEP event shapes \cite{linde:here}
with a larger error quoted to account for theoretical uncertainties.
The question mark for $\bar{s}^2_W$ concerns the scheme-dependence of
the extracted weak angle.  All of the data are in excellent agreement
with the predictions of the standard model.  The $\chi^2$ obtained
when the results of the four experiments for each observable are
combined is typically $\chi^2/df \sim 0.25-1$, which is low, but not
too unreasonable.

{}From these data one can extract the standard model prediction for the
top quark mass.  The results are shown in Table~\ref{tab3}.
\begin{table} \centering \begin{tabular}{|c|c|c|} \hline Data &
$m_t$(GeV) & $\alpha_s$ \\ \hline LEP \protect\cite{nash} & $157^{+25
\;+17}_{-30 \; -20}$ & $0.142 \pm 0.01 \pm 0.002$ \\ LEP
\protect\cite{nash} & $168^{+22 \;+17}_{-26 \; -20}$ & fixed $(0.124
\pm 0.005)$ \\ LEP $+M_W+ \nu N$ \protect\cite{nash} & $149^{+21 \;
+17}_{-23 \; -21}$ & $0.143 \pm 0.01 \pm 0.002$ \\ LEP $+M_W + \nu N$
\protect\cite{nash} & $157^{+20 \; +18}_{-22 \; -21}$ & fixed $(0.124
\pm 0.005$ \\ All \protect\cite{langacker} & $151^{+21 \; +18}_{-23 \;
-14}$ & fixed $(0.124 \pm 0.010)$ \\ \hline \end{tabular}
\caption{Values of $m_t$ obtained from LEP data and combinations of
LEP with other results.  The first uncertainty is experimental and the
second is from the Higgs mass in the range 50 GeV $-$ 1 TeV.  In the
first and third rows $\alpha_s$ is fit to the data, and is constrained
mainly from the hadronic $Z$ width.  The value obtained is slightly
higher than that obtained from the event shapes.  The other rows use a
fixed $\alpha_s$ determined from event shapes. } \label{tab3}
\end{table} The best fit to the data implies $m_t \sim 150$~GeV,
though with large uncertainties.  It is interesting that CDF and D0
should be able to reach $m_t$ values $O(150)$GeV in the next run
\cite{lockyer:here}.  The last row in Table~\ref{tab3} is a global fit
to all $Z$, $W$, and neutral current data assuming $\alpha_s = 0.124
\pm 0.010$.  One also obtains \cite{langacker}
\beq \begin{array}{cl}
\overline{\rm MS}: & \sin^2 \hat{\theta}_W (M_Z) = 0.2325 \pm
0.0007 \\
{\rm on-shell:}& \sin^2 \theta_W \equiv 1 - \frac{M_W^2}{\bar{M}^2_Z} =
0.2257 \pm 0.0026 \end{array} \eeq
for the weak angle in the $\overline{\rm MS}$ and on-shell schemes.
The uncertainties are mainly due to $m_t$; the $\overline{\rm MS}$
definition is considerably less sensitive.  The $m_t$ value in the
last row does not include 2-loop corrections of the form $\alpha
\alpha_s m^2_t$.  It would increase by some 9~GeV if the perturbative
estimate of these terms were included.  However, there is theoretical
uncertainty in the coefficient.  It should be noted that there is no
significant sensitivity to the Higgs mass $M_H$ as long as $m_t$ is
not known independently.

An interesting development is that the four LEP groups have done a
combined study of their results \cite{i2}.  Their basic inputs are the
observables $M_Z$, $ \Gamma_Z$, $\sigma^h_0$, $\Gamma_{l\bar{l}}$,
$g_{V}$, and $g_{A}$.  They found that one obtains essentially the
same result from a joint analysis as from simply averaging the
individual experiments (taking common systematic errors properly into
account).  They also present an average correlation matrix.  This
joint analysis is very useful, and I would like to encourage that it
be continued in the future.

One disturbing aspect of the analysis has been described by Navelet
\cite{navelet}.  Navelet and collaborators have reanalyzed the
infrared divergences associated with virtual photon exchange between
initial and final charged particles.  In order to regulate the
divergences one can take the external fermions off-shell and give the
photon a mass $\mu$.  The physical limit involves returning to $p^2
\ra m^2$ and $\mu \ra 0$.  Navelet argued that these two limits do not
commute, introducing an ambiguity, and that one should take $\mu \ra
0$ first rather than the usual procedure.  He then finds that certain
$\alpha/v$ terms are absent, changing the formulas for the $Z$ widths
by some $O(4\%)$ from the usual formulas.  This a substantial
correction compared to the experimental uncertainties, and it is
crucial that this issue be resolved quickly.

There was considerable discussion of the future LEP program.  Treille
\cite{treille} emphasized that at present the number of events is some
$\sim 300 K/exp$, allowing a precision of $\Delta \sin^2 \theta_W \sim
0.0013$   from the $Z$ widths and asymmetries, which is much less
precise than the value from $M_Z$.  (The comparision of the two is
sensitive to now physics.)  He advocated the importance of
accumulating a few$ \x 10^6/exp$ in the future, yielding an
uncertainty of $0.00065$.  Only at that point will the experiments be
systematics limited.  Mikenberg \cite{mikenberg:here} described the
opportunity for a high luminosity LEP (HLEP) in which there would be
$36 \x 36$ bunches, allowing a luminosity of $(1.5 - 2 ) \x 10^{32}\;
cm^{-2} s^{-1}$.  HLEP could be run at some time after the LEP 200
program.  It would allow a precision of $\Delta \sin^2 \theta_W \sim
0.00035$ from $A_{FB} \, {(b)}$.  It would also be possible to measure
$\Gamma_{b \bar{b}}$ to $ \sim 1\%$ precision.  This would be useful
because $\Gamma_{b\bar{b}}$ receives vertex corrections involving the
top quark, and would allow a separation of the effects of $m_t$ from
other radiative corrections or non-standard Higgs representations
which affect the $W$ and $Z$ masses.

Treille \cite{treille} described the possibility of measuring the
polarization asymmetry, $A_{LR}$, the ``queen of observables'', at LEP.
A measurement $\Delta A_{LR} \sim 0.3\%$ would allow a precision
$\Delta \sin^2 \theta_W \sim 0.0004$.  This would be comparable to the
value obtained from $M_Z$ and would allow a stringent test of the
standard model and search for new physics.


There are a number of practical issues for future LEP 100 running.  To
exploit the high statistics that will be forthcoming it will be
necessary to improve the luminosity measurement to $\Delta \L/\L <
0.1\%$\cite{gurtu,lee}.  Work is under way to improve both the
luminosity monitors themselves and the theoretical calculations that
will be needed to exploit them.  More theoretical work is also needed
is to build new event generators with two or more hard $\gamma$'s in
the final state, both to compare data with the standard model
predictions and to lay the groundwork for searching for new physics in
the $\mu \mu (n\gamma)$ and $q \bar{q} \gamma$
channels\cite{wenniger,demin}.  As was previously mentioned, there is
a theoretical uncertainty
\beq \delta \Delta r \sim \frac{\Delta \alpha (M_Z)}{\alpha (M_Z)}
\sim 0.0009 \eeq
from the low-energy hadronic vacuum polarization.  This is the major
theoretical uncertainty in the relation between $M_Z$ and $\sin^2
\theta_W$, and also in $g_\mu -2$.  New high-precision low energy
measurements are needed \cite{treille}.

As precision gets higher it will be necessary to pay more attention to
higher-order terms in the electroweak predictions.  Closely connected
is the proliferation of definitions of values of $\sin^2\theta_W$.  I
personally get very confused.  At present, most of the uncertainty is
hidden by the experimental errors, but in the future the issue will be
more important.  It would be useful if results were always presented
in terms of the $\overline{\rm MS}$ value $\sin^2 \hat{\theta}_W$
which is useful for comparison with grand unification.  Another
scheme, the on-shell $\sin^2 \theta_W \equiv 1 - M_W^2/M^2_Z$, is also
useful and is easy to translate into the $\overline{\rm MS}$ scheme.
The various effective values that are quoted are difficult to relate
to the rest of the world, and more care in stating the definitions and
translations is necessary in the future.

Work is also needed is to improve the theoretical error in $\alpha_s
(M_Z)$, which has many implications for standard model tests
\cite{linde:here}.  Finally, some caution is needed in the definitions
of $M_Z$ and $\Gamma_Z$.  The forms generally used now are based on
Breit-Wigner formulas, but an alternate definition involves the actual
location of the $Z$ pole.  The relation between the two is under
control but one should be careful.

The LEP 200 machine parameters have never been well-defined.
Possibilities for the energy include $\sqrt{s} = 175, \; 190$, and
$240\;GeV$ and suggested luminosities are in the range $125\;pb^{-1}$
to $ 500\; pb^{-1}$.  It was emphasized by Treille \cite{treille} and
Janot \cite{janot} that the higher energy would be a major advantage
for a number of types of physics.  LEP 200 should make a precise
measurement of the $W$ mass with $\Delta M_W \sim 60$~MeV, and, as has
been described, would allow a search for intermediate-mass standard
model and supersymmetric Higgs particles.

LEP 200 would also allow a measurement of the $\gamma WW$ and $ZWW$
vertices, which would be useful for the text books.  One can, of
course, search for anomalous non-abelian vertices.  However, de Rujula
{\it et al}., \cite{rujula,rujula2} have argued that most previous
estimates of the sensitivity to anomalous couplings were greatly
exaggerated because they did not properly taken gauge invariance into
account.  They argued that any new physics is likely to be electroweak
gauge invariant; otherwise, the successes of the standard model would
be distroyed.  Secondly, they cataloged gauge invariant sets of
operators and claimed that the LEP 100 results have already excluded
virtually any chance for LEP 200 to observe anomalous vertices.  This
idea is probably correct and should be taken seriously\footnote{Some
aspects of the argument have recently been questioned \cite{i50a}.}.
However, by
utilizing such effects as polarizations LEP 200 may be able to place
somewhat better constraints \cite{i3} on the anomalous vertices than
those assumed in \cite{rujula2}.

Kurihara \cite{kurihara} described automated computer calculations of
complicated processes such as $e^+e^- \ra W^+W^- \nu\bar{\nu}$.

\subsubsection{The Weak Neutral Current}

The LEP measurements are extremely precise, but they are blind to
types of physics which don't directly affect the properties of the
$Z$, such as $Z'$ bosons which do not mix with the ordinary $Z$ or new
types of interactions.  A number of other types of precision
observables are therefore important and will be a useful complement to
present and future LEP measurements.  Particularly important are
$M_W$, future deep-inelastic neutrino scattering experiments, and
atomic parity violation.  The $W$ mass will be measured precisely in
several types of experiments: LEP 200 is expected to measure to 60
MeV, the hadron colliders CDF and D0 to about 100 MeV, and HERA to
about 100 MeV.

Enomoto \cite{enomoto} described the TRISTAN program.  He emphasized
that the machine is still running and will be for two or three more
years.  In the past there was a small anomaly in the total hadronic
cross section, $R_{\rm had}$, which was somewhat higher than the
standard model predictions.  However, new calculations of the
radiative corrections have eliminated most of the effect.  New
measurements of the leptonic cross sections and asymmetries, $R_l$,
and $A_l$, in excellent agreement with the standard model were also
presented.

Cocco \cite{cocco} described the latest CHARM~II results.  Previously
they concentrated on $\sin^2 \theta_W$, which can be cleanly measured
from the ratio of neutrino and antineutrino scatterings.  They have
now extracted the individual $\nu_\mu (\bar{\nu}_\mu) e^-$ elastic
scattering cross sections, from which they are able to determine the
vector and axial couplings, $g^e_{V,A}$, relevant to the four-fermi
neutrino-electron interaction.  In the standard model to lowest order
these are the same as the vector and axial vector couplings of the $Z$
to the electron, $g_{V,A}$, that are measured at LEP.  However, if
there is new physics they are not quite the same, so it is important
to measure them in both ways.  CHARM~II obtained
\begin{eqnarray} g^e_V &=& - 0.025 \pm 0.014 \pm 0.014 \nonumber \\
                 g^e_A &=& - 0.503 \pm 0.007 \pm 0.016,\end{eqnarray}
which are in agreement with the standard model values $-0.037 \pm
0.001$ and $0.506 \pm 0.001$.

Bolton \cite{bolton} described new measurements of deep inelastic
neutrinos scattering by the CCFRW group at Fermilab.  They have
extracted the on-shell value $\sin^2 \theta_W \equiv 1-
\frac{M_W^2}{M_Z^2}$, which is insensitive to $m_t$ for $\nu N$
scattering. They obtained
\beq             {\rm CCFRW} \;\;\;0.2242 \pm 0.0042 \pm [0.0047] \eeq
which is comparable in precision to the CDHS and CHARM measurements
\begin{eqnarray}
{\rm CDHS} && 0.228 \pm 0.005 \pm [0.005] \nonumber \\
{\rm CHARM} && 0.236 \pm 0.005 \pm [0.005]. \end{eqnarray}
The second uncertainty is theoretical, mainly associated with the
$c$-quark threshold.  In the future they expect to improve their
experiment significantly.  One of their major systematic problems,
$\nu_e$ contamination of the beam, should be reduced considerably by a
new sign-selected quadrupole beam.  The uncertainty in the $c$-quark
threshold will be reduced by measuring both $\nu_\mu$ and
$\bar{\nu}_\mu$; appropriate combinations will reduce the sensitivity.
They expect to achieve $\Delta \sin^2 \theta_W \sim 0.0025$ even
without the new Fermilab main injecter, and 0.0015 with it, including
all experimental and theoretical uncertainties.  There was also a
description of a careful study of backgrounds in the CCFRW experiment
\cite{sandler}.  This removed essentially all of the anomalous
same-sign dimuons which had apparently been present for a long time in
many of the neutrino experiments.  There no longer seems to be a
significant problem.

A number of other future weak neutral current experiments, especially
atomic parity violation, were described by Treille \cite{treille}.

\subsubsection{Searches/Parametrizations for New Physics}

There are a number of ways to  parametrize data to maximize
sensitivity to new physics.  These are complementary, and each has its
advantages.

\begin{itemize}

\item One possibility is to study generic models, such as the effects
of additional heavy $Z'$ bosons, or the mixing of the ordinary with
exotic fermions such as $d_L \leftrightarrow D_L$.  The current limits
on the $Z'$ in $E_6$ models are shown in Table~\ref{tab4}.  Other $Z'$
models were described by Kneur \cite{kneur} and Casalbuoni
\cite{casalbuoni}.
\begin{table} \centering \begin{tabular}{|lrrrr|}
\hline
           & $Z_\chi$  & $Z_\psi$ & $Z_{\rm LR}$ & $Z_n$ \\ \hline
Direct CDF & 280       & 180      & 240          & 240   \\
LEP $+$ WNC & 320      & 160      & 390          & 180   \\
\ \ \ \ (mixing arbitrary)& & & & \\
LEP $+$ WNC & 550 & 160 & 860 & 210 \\
\ \ \ \ (mixing constrained) & & & & \\ \hline
\end{tabular}
\caption{Limits on the masses of various $Z'$ bosons which occur in
the $E_6$ model, in GeV.  The        limits from CDF in the first
column are
from direct searches \protect\cite{maeshima:here}. The other rows are
indirect limits from precision experiments, with    the $Z Z'$ mixing
   arbitrary or constrained in specific models \protect\cite{i4}.}
\label{tab4}
\end{table}

\item One can also consider specific models; for example, in which
there are $Z'$s, Higgs representations, and exotic fermions with their
properties correlated.  These are less general in terms of the
properties of the $Z'$, for example, but show useful correlations
between the effects of the types of new physics.

\item Effective operators are another possibility.  In many cases
there are too many of these to be useful.  However, in the case of
showing the constraints of LEP 100 and how they affect new physics at
LEP 200 they are extremely useful \cite{rujula,rujula2}.

\item Another possibility are the $S, \; T, \; U$ parameters.  These
apply to all observables, but, by definition, only describe types of
new physics which only affect the $W$ and $Z$ self-energies.

\item Altarelli \cite{altarelli} described an alternate formalism
based on three parameters $\epsilon_1, \; \epsilon_2,\; \epsilon_3$.
Theses are defined in terms of the deviations of the three observables
$M_W/M_Z$, $\Gamma_{l\bar{l}}$, and $A_{FB}(l)$ from the standard
model predictions.  These are more general than $S$, $T$, and $U$ in
the sense that they can parametrize any type of new physics.  However,
they have the shortcoming that one cannot extend this parametrization
to other observables unless additional assumptions are made.

\item Finally, one can define the $\rm a^{\rm th}$ component of the
deviation vector \cite{LLM} $(O_a - O_a^{SM} (M_Z))/\Delta O_a \equiv
V_a$, which is the deviation of the $\rm a^{\rm th}$ observable from
its standard model prediction normalized by the total (experimental
$+$ theoretical) uncertainty in the measurement.  The deviation vector
would be most useful if deviations are actually seen.  Its direction
(length) is characteristic of the type (strength) of the new physics.

\end{itemize}

\subsubsection{1 TeV Scale: Astrophysics/Cosmology}

Another probe of new physics at the TeV scale involves astrophysics
and cosmology.  There was an interesting talk by Freeman
\cite{freeman}
on Galactic dark matter, which             emphasized the observations
of the density,    evidence for dark matter     in the Galaxy, and
possible interpretations.  The densities, relative to the critical
density, of matter on various scales are summarized in
Table~\ref{tab5}.
\begin{table} \centering
\begin{tabular}{|ll|} \hline
$\Omega_{\rm visible} \sim 0.007$ & -- -- \\
$\Omega_{\rm halo} \sim 0.07$ & HI rotation curves \\
$\Omega_{\rm clusters} \sim 0.1-0.3$ & -- -- \\
$\Omega_{\rm baryon} \sim 0.02-0.1$ & nucleosynthesis \\ \hline
\end{tabular}
\caption{Values of the density (relative to the critical density) of
visible matter, of the matter clustered on the scale of halos and
clusters, and the baryon density     inferred from
nucleosynthesis \protect\cite{freeman}.}
\label{tab5}
\end{table}
{}From this we see that baryons in some form could account for the dark
matter in halos, {\it i.e.,} $\Omega_{\rm halo}$, consistent with the
normal nucleosynthesis scenario.  There are two experiments, the MACHO
(Massive Astrophysical Compact Halo Objects) and Saclay experiments,
both of which look for gravitational microlensing of distant stars to
search for small objects that could comprise baryonic dark matter.
They should be able to observe objects in the entire relevant mass
range $(10^{-8} - 10^{+2}) M_\odot$.  If these experiments see no
effect one could essentially rule out baryonic dark matter, leaving
the possibilities of WIMPS or massive neutrinos.

\subsubsection{The TeV Scale: CP Violation}

CP violation is strongly suppressed in the standard model and is
therefore an excellent place to look for new physics.  So far the only
indication of CP violation is in the kaon system, in which one
observes the two parameters $\epsilon$ and $\epsilon'$. $\epsilon$ can
be generated by $K_1 - K_2$ mixing (indirect CP violation) as well as
by direct CP-breaking in the decay amplitudes. $\epsilon'$ can be
generated only by direct CP breaking.  One expects $\epsilon'/\epsilon
\neq 0$ in the standard model due to phases in the CKM matrix.  The
gluon penguin diagrams yield $\epsilon'/\epsilon \sim$ few$\x 10^{-3}$.
However, above $m_t = 100 \; GeV$ additional electroweak penguins
which can cancel the gluon effects are important, and there are also
complications from isospin breaking due to the quark masses $m_d \neq
m_u$.  All of these effects have theoretical uncertainties, and they
can cancel, so the prediction is very uncertain.  For large $m_t$ one
expects smaller values of $\epsilon'/\epsilon \neq 0$, and it could go
through zero, {\it e.g.}, at 200 GeV.

The experimental situation is equally confused.  For years there has
been a discrepancy between the Fermilab and CERN experiments.  Barker
\cite{barker} presented a new preliminary value from FNAL E731:
\beq  R e \frac{\epsilon'}{\epsilon} = \left[ 6.0 \pm 5.8 \pm 3.2 \pm
1.8\right] \x 10^{-4}.\label{eq1} \eeq
The central value is now positive but it is still consistent with
zero.  Final results from E731 are expected very soon.  One
anticipates that the final uncertainty will be $\left[\pm 5.1 \pm 3.2
\right] \x 10^{-4}$.  (The last error in (\ref{eq1}) is due to Monte
Carlo uncertainties, which should be eliminated in the final value.)
The future  Fermilab E832 experiment should yield a precision of
$10^{-4}$.  New results from the CERN experiment NA31
(Perdereau \cite{perdereau})
based on 1989 data still indicate a positive and non-zero value:
\begin{eqnarray} {\rm 1989 \;\; (preliminary):} && (2.1 \pm 0.9) \x
10^{-3} \nonumber \\ {\rm 1986-89 \;\;(preliminary):} && (2.3  \pm
0.7) \x 10^{-3}. \end{eqnarray}
There is a mild discrepancy between the two experiments and until this
is resolved it is not clear what is going on.  The future CERN
experiment NA48 will also have a precision of $10^{-4}$.  The CPLEAR
group \cite{fry} will measure to a precision of $(2 - 3) \x 10^{-3}$,
and will also measure a number of other quantities such as CPT phases.
By around 1998, the DA$\Phi$NE \cite{leefranzini} $\phi$ factory will
measure $\epsilon'/\epsilon$ to a precision of $10^{-4}$, and many
other CP and CPT observables simultaneously and precisely.

Another way of probing CP breaking in the kaon system is the rare
decay $K_L \ra \pi^0 e^+e^-$ \cite{ohl}.  The dominant decay is via
$K_L \ra \pi^0 \gamma^*$, followed by $\gamma^* \ra e^+e^-$, which is
CP-violating.  One expects a contribution of less than $2 \x 10^{-12}$
from the indirect $K_1 - K_2$ mixing \cite{zeller}.  The more
interesting direct mechanism due to electroweak penguins and $WW$
boxes, {\it etc.}, is expected to yield a branching ratio around
$10^{-11} - 10^{-12}$.  There is also a CP-conserving contribution via
the two-photon intermediate state $K_L \ra \pi^0 \gamma^* \gamma^*$,
$\gamma^* \gamma^* \ra e^+e^-$, which is strongly suppressed by
$\alpha$.  There was considerable theoretical controversy as to
whether this would be large enough to be serious.  However,
measurements by NA31 \cite{heinz} now indicate that this contribution
is less than $4.5 \x 10^{-13}$ and therefore not a problem.  However,
there is a very serious background \cite{greenlee} from $K_L \ra
e^+e^- \gamma\gamma_{\rm brems}$, which yields a contribution of order
$7 \x 10^{-11}$.  The extent to which the signal can be separated from
this background depends on the resolution, but it may prove fatal.

A number of rare $K$ decays which are being searched for mainly at
Brookhaven and KEK, are listed in Table~\ref{tab6} along with their
current limits, their expectations, and why they are interesting.
\begin{table} \centering \large \def\baselinestretch{.95}  \small
\begin{tabular}{|p{2.8cm}p{3.0cm}p{2.3cm}p{2.5cm}p{3.0cm}|} \hline
Mode & Limit/Value 90\% & Future & SM & Why Important \\ \hline
$K_L \ra \pi^0 e^+e^-$ & $ <      5.5 \x 10^{-9}$ \protect\cite{ohl} &
$1 \x 10^{-11}\protect\cite{barker799}  \newline
[7 \x 10^{-11} bkg]$ &
$10^{-11} - 10^{-12}$ & direct CP \newline $K_L \ra \pi^0 \gamma \ra
\pi^0 e^+e^-$ \\
$K_L \ra e^+e^- \gamma\gamma_{br}$ & $6.6 \pm 3.2 \x
10^{-7}$ \protect \cite{ohl} & & $5.8 \x 10^{-7},$ \newline
$E^*>5$MeV & bkg to $\pi^0 e^+e^- $ \\
$K_L\ra \pi^0 \gamma \gamma $ \newline
$+$ Dalitz plot & $1.7 \pm 0.3 \x
10^{-6}$\protect \cite{heinz} & & ChPT $0.7 \x 10^{-6} $ & CP even
$K_L \ra \pi^0 \gamma\gamma$ $\ra \pi^0 e^+e^-$ $<4.5 \x 10^{-13}$ \\
$K^+ \ra \pi^+ e^+e^-$ \newline $+$  spectrum & $2.75 \pm 0.26 \x
10^{-7}$\protect \cite{zeller} & && indirect cont to $K_L \ra \pi^0
e^+e^-$ $< 2 \x 10^{-12}$ \\
$K_L \ra \pi^0 \nu \bar{\nu}$ & $< 2.2 \x
10^{-4}$\protect \cite{barker} & $10^{-8}$ & $10^{-10} - 10^{-12}$ &
CP \\
$K^+\ra \pi^+ \nu \bar{\nu}$ & $ <       5 \x
10^{-9}$\protect \cite{ito} & $ 2 \x 10^{-10}$ \newline
(AGS booster) & $(1-6)
\x 10^{-10}$ & $V_{td}$ \\
$K_L \ra e^+e^- \gamma$ & $9.1 \pm 0.6 \x 10^{-6}$ \protect\cite{ohl}
& & $9.6 \pm 0.4 \x 10^{-6}$ & ChPT \\
$K_L \ra \mu^+ \mu^-$ & $7.9 \pm 0.6 \pm 0.3 \x
10^{-9}$ \protect\cite{komatsubara} \newline $6.96 \pm 0.40\pm 0.22
\x 10^{-9}$ \protect\cite{BNLE791} & & $> 6.8 \pm 0.3 \x 10^{-9}$ &
CPT + Unitarity; $m_t$ (?) \\
$K_L \ra e^+e^-$ & $< 1.6 \x 10^{-10}$ \protect\cite{komatsubara}
\newline $<5
\x 10^{-11}$ \protect\cite{BNLE791} & $8.5 \x 10^{-13}$ & $10^{-12}$ &
\\ %
$K_L \ra \mu e$ & $<9.4 \x 10^{-11}$ \protect\cite{komatsubara}
\newline
$<3.3 \x 10^{-11}$\protect \cite{BNLE791} & $2 \x 10^{-12}$ & 0 & FCNC
in ``non-standard non-standard models'' \\
$K^+ \ra \pi^+ \mu^+ e^-$ & $<2.1 \x 10^{-10}$\protect\cite{zeller} &
$10^{-12}$ & 0 & \\    \hline
\end{tabular}
\caption{Some rare decay modes, their current limits or values,
standard model expectations, and why they are particulary useful.}
\label{tab6}
\end{table}

Other related searches for CP breaking include the electric dipole
moments of the neutron and of atoms.  Barr \cite{barr} described the
dipole moments of atoms, which can be generated, amongst other things,
by the electron dipole moment $d_e$ and by anomalous $eN$
interactions.  These may yield measurable effects in extended models,
in which CP violation may be mediated by Higgs exchange.  The
expectations are much smaller in the CKM model.  Eilam \cite{eilam}
discussed the possibility of CP breaking in asymmetric $t$ decays.
These are much too small to be observed in the standard model, but
could be important if there are some types of exotic new physics.

\subsubsection{CP Violation in the $B$ System}

The holy grail of CP breaking is the $B$ system.  One expects large
effects because of enhanced CP-violating phases and because
ordinary $B$ decays are strongly suppressed.  The general situation
was described by Bigi \cite{bigi}, and other talks dealt with $B$
physics at hadron colliders and the fact that a dedicated experiment
at Fermilab would effectively be a $B$ factory \cite{lockyer:here},
the HLEP option \cite{mikenberg:here}, CP violation in the $B$ system
\cite{nakada}, rare decays \cite{haba}, and machine possibilities
\cite{baconnier} for asymmetric $(e^+e^-)  B$ factories.

To observe and interpret CP violation in the $B$ system and thus to
stringently test the standard model one needs higher rates for $B$.
The optimal strategy, {\it i.e}., to use hadron colliders, HLEP, or an
asymmetric $B$ factory, is still not clear.  In addition to the
machines, there is much background work necessary to develop the
theoretical knowledge to interpret the results.  Before and during the
$B$ factories a full program of studies of $B$ decays will be needed
to make reliable phenomenological models.

The $B_s$ mixing is predicted to be nearly maximal in the standard
model, which makes it very hard to measure.  It is important to verify
this.  The ratio $|V_{ub}/V_{cb}|$ is important.  A new measurement
\beq \left| \frac{V_{ub}}{V_{cb}} \right| = (9.4 \pm 1.0 \pm 0.8) \%,
\eeq
which uses the whole spectrum, was reported from ARGUS
\protect\cite{paklovhere}.  The measurements are now very good, but we
still need better theoretical models to reduce the theoretical
uncertainty in the extraction of $|V_{ub}/V_{cb}|$.  We also need
better calculations for $B_{d,s} \leftrightarrow \bar{B}_{d,s}$ mixing
as well as the value of $m_t$ to extract $V_{td}$ reliably.  An
alternative and complementary way to extract $V_{td}$ is from the rare
decay $K^+ \ra \pi^+ \nu \bar{\nu}$.

The goal is to construct and overconstrain the unitarity triangle
shown in Figure~\ref{fig1}.
\begin{figure}
\vspace{7cm}
\caption{The unitarity triangle.  The lengths of the sides are
magnitudes of elements of the CKM matrix.  The angles are CP-violating
phases.}
\label{fig1}
\end{figure}
Testing whether the vectors really add up to a triangle probes such
new physics as a heavy $W_R$ boson, fourth-family fermions, and new
sources of CP violation.  The sides of the unitarity triangle are
magnitudes of the CKM elements.  Two can be constructed from the
partial rates for semi-leptonic $B$ decays into $c$ and $u$ quarks,
while $|V_{td}|$ can be extracted from $B^0_d \leftrightarrow
\bar{B}^0_d$ or $K^+ \ra \pi^+ \nu \bar{\nu}$, provided one knows
$m_t$.  Bigi \cite{bigi} emphasized that one of the CP-violating
angles, $\varphi_1$, can be determined from CP breaking in the kaon
system, namely from the ratio $|\epsilon_\kappa|/\Delta m_{B_d}$.
Independent measurements of $\phi_1$ and the other angles can be
obtained from CP asymmetries in the $B$ system, expecially the
time-dependent asymmetries
\beq A(t) = \frac{\Gamma (B \ra f)_t - \Gamma( \bar{B} \ra
\bar{f})_t}{ \Gamma (B \ra f)_t + \Gamma(\bar{B} \ra \bar{f})_t}.\eeq
The cleanest determinations theoretically can be made in the case that
$f$ is its own CP conjugate, $f = \bar{f}$.  Interferences between the
phases in the mixing and the decay amplitude lead to CP asymmetries if
there is only one process contributing to the decay.  For example, one
measures the three angles of the triangle from the typical decays
\begin{eqnarray} B_d \ra \psi K_s &,& \sin 2 \varphi_1 \nonumber \\
                 B_d \ra \pi^+\pi^-&,& \sin 2 \varphi_2 \nonumber \\
         B_s \ra K_s \rho_0 &, & \sin 2 \varphi_3.\end{eqnarray}
The first is very clean.  The others suffer from possible penguin
pollution: penguin as well as tree diagrams are both present, leading
to some theoretical uncertainty.

\subsubsection{Weak Scale Baryogenesis}

It has long been known that $B$ and $L$ are violated by anomalies in
the standard model \cite{hooft},
\beq \partial \cdot J_B = \partial \cdot J_L = - \frac{3g^2}{16\pi^2}
Tr F \tilde{F}. \eeq
This can be thought of as tunneling between vacua of different $B+L$
(Figure~\ref{fig2}).
\begin{figure}
\vspace{6cm}
\caption{Schematic of baryon number violation in the standard model.
The minima correspond to degenerate vacua of different $B+L$.  The
anomaly describes tunneling between these vacua.}
\label{fig2}
\end{figure}
The anomaly conserves $B-L$.  At zero temperature the effect is
irrelevant for practical purposes because
the tunneling rate is suppressed by the factor
$\exp(-4\pi/\alpha_W) \sim 10^{-170}$.  However, for temperatures
comparable to the electroweak scale, {\it i.e.,} 1 TeV, there may be
unsuppressed thermal fluctuations.  A specific solution that describes
these transitions is known as the sphaleron.  One serious consequence
is that any baryon asymmetry of the universe produced earlier, such as
in a GUT epoch, would be washed out by the electroweak $B+L$
violation, unless the initial asymmetry had a non-zero $B-L$ or the
initial $B+L$ asymmetry was huge, {\it i.e.,} of $O(1)$
\cite{shaposhnikov}.  If the baryon asymmetry was washed out a new
asymmetry
\beq \frac{n_B - n_{\bar{B}}}{s_\gamma} \sim 10^{-11} \eeq
must have been created at the time of the electroweak transition.
This is possible in principle but difficult in practice.  It is hard
to achieve sufficient CP violation within the standard model; one must
add new physics to enhance the CP breaking.  The necessity for being
out of equilibrium can be associated with expanding bubbles of true
vacuum in a first order phase transition.  There a number of
possibilities for the mechanism, such as reflection from the wall of
the expanding bubble \cite{cohen}.

A related possibility is baryon number violation in high energy hadron
collisions, of order $\sqrt{s} \simgr 10$~TeV, such as the SSC.  It
has been speculated that such collisions might produce $B$-violating
transitions.  These would manifest themselves not only by the $B$
violation but, more dramatically, by the production of $O(100)W$ and
$Z$ particles associated with the fields that characterize the
different vacua.  The cross section is
\beq \sigma_{\rm tot} \sim \exp \left( \frac{4\pi}{\alpha_W} F
(E/E_{\rm crit} ) \right), \eeq
where
\beq F(\epsilon) = -1 + \frac{9}{8} \epsilon^{4/3} - \frac{9}{16}
\epsilon^2 + \cdots \eeq
The effect will be strongly suppressed if $F<0$, and the results are
unphysical (nonunitary) for $F>0$.  However, if $F=0$ there would be
huge effects.  The calculation of $F$ is extremely difficult because
perturbation theory does not hold in the relevant domain; this is an
open topic of debate amongst the theorists.  This situation was
reviewed by Ringwald \cite{ringwald}, who expressed the hope that we
would have a reliable answer within the next two years.  Novikov
\cite{novikov} argued that a previous calculation in a supersymmetric
theory is not valid.

\subsection{100 TeV: Flavor Changing Neutral Currents}

Flavor changing neutral currents (FCNC) in such decays as $\mu$,
$\tau$, $K$, $B$, $\cdots$ are an important manifestation of physics
at the 100~TeV scale.  Searches for such decays place stringent
constraints on dynamical symmetry breaking, family symmetries,
extended Higgs sectors, compositeness, and leptoquarks, all of which
are expected to mediate effects.  At this meeting Zeller \cite{zeller}
presented results
\begin{eqnarray} K^+ \ra \pi^+ \mu^+ e^- && B < 2.1 \x 10^{-10}
\nonumber \\
 K^+ \ra  \mu^{\pm} e^{\mp} && B< 3.3 \x 10^{-11} \end{eqnarray}
from the BNL experiments E777 and E791.  These are very impressive.
However, most types of physics which would lead to these particular
decays are already strongly constrained by the $K_L-K_S$ mass
difference, and any observable effects would have to be due to
``non-standard non-standard'' physics.  Hou \cite{hou} described the
possibility of $t \ra cH$ or $H \ra \bar{t}c$ in models for which
there is no natural flavor conservation in the Higgs sector, and
Fritzsch \cite{fritzsch} emphasized the decay $\mu \ra 3e$, which can
occur by mixing between right-handed singlets and doublets.  In most
models the FCNC are largest for the third family.

\subsection{$10^{2-19}$ GeV: Neutrino Mass}

Many extensions of the standard model predict non-zero neutrino mass
at some level.  Typically one expects $m_\nu \sim v^2/M \ll v$, where
$v$ is the weak scale and $M$ is the scale of new physics.  Therefore,
small neutrino masses probe large-scale physics; and there are
                     interesting predictions in various models for
grand unification \cite{plangacker}.  There is some     evidence for
neutrino
masses from Solar neutrinos, atmospheric neutrinos, the possible 17
keV neutrino, and hot dark
matter \cite{plangacker,litchfield,morrison}.

\subsection{$10^{16} - 10^{19}$ GeV: The Ultimate Unification}

One of the great dreams is a unification of the fundamental
interactions into a simpler structure, perhaps even with gravity.  The
old fashioned (and perhaps naive) view of grand unification was that
the standard model should be embedded in a simple group $G$ at a scale
$M_X \sim 10^{14 - 16}\;GeV$.  This is sufficiently below the Planck
scale, $M_P \simeq 10^{19}\;GeV$, that it perhaps makes sense to
ignore gravity in this partial unification.  In the old-fashioned
early days of GUTS, model builders typically invoked very large Higgs
representations for whatever purpose they desired.  The modern
string-inspired view \cite{louis} is that there should be a direct
unification of all of the forces at the string compactification scale
$10^{18} - 10^{19}\;GeV$.  It is unlikely, through not impossible,
that there is an isolated GUT below the $10^{18}\;GeV$ scale.  Even if
one has such a situation, it is unlikely that one would have large
Higgs representations.  Despite these prejudices, there is
experimental evidence, using precise LEP data on the low energy
couplings, that the three (properly normalized) gauge couplings
$\alpha_3$, $\alpha_2$, $\alpha_1$ meet at a point when extrapolated
in the minimal supersymmetric extension of the standard model (MSSM),
at a unification scale of around $M_X \sim 2 \x 10^{16}\;GeV$.  (See
Figure \ref{fig3}.)  This could well be an accident, but could also be
a hint that the bang scenario and supersymmetry may be correct.  One
should not take the details too seriously because of the many
theoretical uncertainties, but perhaps some sort of SUSY-unification,
with or without strings, is relevant.
\begin{figure}
\vspace{15cm}
\begin{caption}\ \ Extrapolation of the normalized gauge couplings in the
standard model and its supersymmetric extension using
$ \alpha^{-1} (M_Z)  =  127.9 \pm 0.2$, $
\sin^2 \hat{\theta}_W (M_Z)  =   0.2325 \pm 0.0007,$ and
$ \alpha_s (M_Z)  =  0.124 \pm 0.010.$
Clearly the standard model does not unify into an ordinary GUT,
whereas the supersymmetric extension is compatible with grand
unification.  Ordinary GUTS are also excluded by the non-observation
of proton decay, while in SUSY GUTs proton decay via the ordinary
$d=6$ operators is strongly suppressed.  However, $d=5$ operators are
still dangerous.
\end{caption}
\label{fig3}
\end{figure}
Another way of seeing this is that  one can predict $\alpha_s$ from
the observed values of $\alpha$ and $\sin^2 \theta_W$:
\begin{eqnarray} {\rm SM:} && \alpha_s (M_Z) \sim 0.072 \pm 0.001 \pm
0.01 \pm ? \nonumber \\
{\rm MSSM:} && \alpha_s (M_Z) \sim 0.120 \pm 0.003 \pm 0.004 \pm 0.01
\pm ? .\end{eqnarray}
The first uncertainty is from the input parameters.  The second in the
supersymmetric case is from the masses of the superpartners, and the
next is from the splittings of the superheavy particles.  The question
marks are the possible effects of adding new multiplets that are split
into light and heavy sectors.  A comparison with the experimental
values in Table~\ref{tab1} shows the success of the MSSM.

There are a number of interesting theoretical and experimental
implications of unification.
\begin{itemize}
\item One is the possible connection with supersting theories.
\item Real unified theories that exist as a separate gauge group
predict proton decay at some level.  Experimentally, the limit on two
important modes are $\tau_{e^+ \pi^0} > 10^{33}$~yr and
$\tau_{\bar{\nu} K^+} > 10^{32}$~yr \cite{litchfield}.  The $e^+
\pi^0$ decay excludes the ordinary $SU_5$ model, but is strongly
suppressed in the supersymmetric models due to the larger unification
scale.  In supersymmetric GUTs there is a new mechanism
(dimension-five operators) for proton decay.
\begin{figure}
\vspace{5cm}
\caption{The dimension-five contributions to proton decay.  The
interaction at the right, mediated by the exchange of superheavy
colored Higgsino, must be dressed by the exchange of a wino or other
particle to generate an operator involving only quarks and leptons.}
\label{fig4}
\end{figure}
These are generated by the diagrams in Figure~\ref{fig4} and lead
mainly to decay modes such as $p\ra \bar{\nu} K^+$.  Because the basic
exchange is a fermion the lifetime is $\tau_p \sim M_{\tilde{H}}^2
\sim M_X^2$, which is much more dangerous than the normal diagrams
generated by a boson exchange $(\tau_p \sim M^4_X)$.  The actual decay
rate depends on the details of the spectrum of the superpartners.
Nath and Arnowitt \cite{nath} have made a detailed calculation of the
spectrum in supergravity models consistent with unification.  They
find that the no-scale models of supergravity are excluded by proton
decay, but generalized models have some allowed regions in parameter
space.  These regions correspond to $m_{\tilde{g}} < m_{\tilde{q}}$, a
light Higgs scalar $m_h < M_Z$, $m_t < 175\;GeV$, and a chargino and
two neutralinos with masses
$\simles 100\;GeV$.  Models such as superstring
theories or flipped-$SU_5$ that are not real grand unified theories
may not have any proton decay.

Raby \cite{raby} described some possibilities for the low energy
fermion spectrum in unified theories.  There are too many parameters
to predict anything from first principles, so additional assumptions
are needed.  Dimopoulos, Hall, and Raby have revived an old ansatz due
to Georgi and Jarlskog for the fermion mass matrices at the high
scale, and then run them to low energies.  (In the normal grand
unified theories the ansatz corresponds to the ``bad'' GUTs with large
Higgs representations and discrete symmetries.)  They choose as input
parameters the $e$, $\mu$, $\tau$, $u/d$, $c$, and $b$ masses and
$|V_{cd}|,$ $|V_{cb}|$.  As outputs they generate $m_d$, $m_s$, $m_t
= 180 \pm 10\;GeV$, and $|V_{ub}/V_{cb}| \sim 0.05$.  The latter
prediction is somewhat low compared to experiment.

\item There may also be implications for neutrino masses, and some
string-motivated supersymmetric models can generate masses in
agreement with what is suggested by the Solar neutrino
problem \cite{plangacker}.

\item There may also be implications for cosmology and  the baryon
asymmetry of the universe, at least if $B-L$ is violated so as to
survive the electroweak phase transition.
\end{itemize}

There are other special models that lead to other forms of baryon
number violation \cite{zoupanos} or neutron oscillations $n \ra
\bar{n}$ \cite{gibin}.  These are not the canonical grand unified
theories and do not have any strong motivations.

\section{Conclusion}

There is no evidence for any deviation from the standard model.
However, its many shortcomings suggest that there must be new physics.
The models fall into two broad categories.  The whimper models, in
which there are many scales of physics, are nonperturbative and
include such ideas as dynamical symmetry breaking and compositeness.
The other extreme possibility is the bang scenario, which is
perturbative and in which there are not many new thresholds between
present energies and the Planck scale.  This may be associated with
elementary Higgs fields, supersymmetry, and grand unification.  There
is a hint of support for the bang scenario from the unification of
coupling constants in the MSSM.

We do not know what is the ultimate source of new physics.  There are
many complementary probes, all of which are important and should be
pursued.  These include searches at the large colliders, carrying out
the full program of precision experiments, searches for neutrino mass,
cosmology and the large-scale structure of the universe, and trying to
forge connections with fundamental theories such as superstrings.

\addtolength{\itemsep}{-4pt}

\end{document}